# Laser frequency offset locking via tripod-type electromagnetically induced transparency

Kang Ying,[1] Yueping Niu,[2,3] Dijun Chen,[1,*] Haiwen Cai,[1] Ronghui Qu,[1,4] and Shangqing Gong[2,5]

[1]*Shanghai Key Laboratory of All Solid-State Laser and Applied Techniques,*
*Shanghai Institute of Optics and Fine Mechanics,*
*Chinese Academy of Sciences,Shanghai 201800, China*
[2]*Department of Physics, East China University of Science and Technology,Shanghai 200237, China*
[3]*niuyp@ecust.edu.cn*
[4]*rhqu@siom.ac.cn*
[5]*sqgong@ecust.edu.cn*

We have demonstrated the laser frequency offset locking via the $Rb^{87}$ tripod-type double dark resonances electromagnetically induced transparency (EIT) system. The influence of coupling fields' power and detuning on the tripod-type EIT profile is detailed studied. In a wide coupling fields' detuning range, the narrower EIT dip has an ultranarrow linewidth of ∼590 kHz, which is about one order narrower than the natural linewidth of $Rb^{87}$. Without the additional frequency stabilization of the coupling lasers, we achieve the relative frequency fluctuation of 60 kHz in a long time of ∼2000 s, which is narrower than the short-time linewidth of each individual laser.

## 1. Introduction

Electromagnetically induced transparency (EIT) is a quantum interference phenomenon occurring when two electromagnetic fields resonantly excite two different transitions sharing a common state. It has attracted considerable attention because of its potential applications in many fields [1–15], such as the nonlinear optics, lasing without inversion, the resonant enhancement of the refractive index, atom interferometry, etc. Though the physical mechanism behind EIT has been studied in details throughout the last decade, several potential aspects still remain untouched or rarely addressed to date. One such application is using EIT signal as a frequency discriminator to lock the relative frequency of two lasers. Two lasers with a fixed frequency difference related to atomic energy level splitting are required in many laser-atoms interaction study such as laser cooling and trapping of alkalis, Raman sideband cooling, slow light, atomic clock, etc [16–18]. In some situations, two separated lasers can be used, which are individually referenced to atomic transitions [19–23] or to Fabry-Pérot etalon fringes [24–26]. While, in many experiments, only the frequency difference must be controlled to high precision, but the absolute frequency is less important. In this case, the two frequency components needed for experiments can be generated from a single laser beam with an acoustic optical or electric optical modulator [27]. Alternatively, we can use two lasers and the frequency of one laser can be offset locked relative to the other laser using the interference beat between the two lasers [28]. However, there are some defects in the above mentioned methods. One is the need of some electronics such as detectors, oscillators and associated high-frequency electronics, operating at the offset frequency typically in the microwave, which can be expensive and complex. The other one is that the frequency difference for these methods is not intrinsic to the atomic system of interest, which would produce extra frequency deviation to the laser-atoms interaction study.

The EIT resonance occurs where the frequency difference between the probe and coupling fields precisely matches the energy splitting. So, it provides a direct frequency reference for offset locking the relative frequency of two lasers. Due to its quantum interference origin, EIT resonance spectrum can be much narrower than the usual atom transitions spectrum, with greater dispersion and the potential for lower relative frequency uncertainty. Many studies have been done to offset lock one laser frequency to another using the EIT dip signal since its early demonstration in [29]. S. C. Bell and co-workers have offset locked the lasers frequency via Lambda-type EIT and achieved less than 1 kHz spec-tral width of microwave beat frequency via high bandwidth feedback [30]. Then, laser relative frequency offset locking to exited state transitions via cascade-type EIT has been demonstrated and the combined linewidth of 280 kHz in a short-time has been obtained [31]. All the above studies of frequency offset locking via EIT are based on single dark resonance system, where the width of EIT dip is remarkably dependent on the detuning of the coupling field. When the coupling laser is detuned, the width of EIT dip becomes broader, which degrades the frequency offset locking effect. Therefore, in those single dark resonance systems, stabilizing coupling field frequency to the related atomic transition is essential.

---

* Corresponding author: djchen@siom.ac.cn

In our previous study, we have theoretically investigated an ultranarrow linewidth in a double dark resonances tripod-type EIT system [32]. Theoretical result shows that, compared to the single dark resonance system, the interacting dark states lead to a narrower transparency window and the dispersion in the narrower transparency window can be one order larger than that of the single dark resonance system. Very recently, we have reported an experimental study of the double dark resonances EIT in the tripod-type Zeeman splitting sublevels of $Rb^{87}$ [33]. An EIT dip with linewidth one order narrower than that of Lambda-type EIT has been observed at room temperature. Furthermore, both the theoretical and experimental result have predicted that the ultranarrow spectrum would not be broadened in a wide detuning range of the coupling fields. So, it brings benefit to the frequency offset locking application as a better offset locking effect can be expected without the additional stabilization of the coupling fields' frequency.

In this paper, the narrower EIT dip in the tripod-type double dark resonances $Rb^{87}$ system is used as a reference signal to offset lock the relative frequency of the probe laser and coupling laser. The influence of laser fields' power and detuning on the double dark resonances EIT profile is detailed studied. In a wide coupling fields' detuning range, the narrower EIT dip has an ultranarrow linewidth of ∼600 kHz, which is about one order narrower than the natural linewidth of $Rb^{87}$. With a high speed and a low speed feedback channel, the relative frequency of probe laser and coupling laser are locked. Without the additional frequency stabilization of the coupling lasers, we achieve the relative frequency fluctuation of 60 kHz in a long time of ∼2000 s, which is much narrower than the short-time linewidth of each individual laser.

## 2. Experiment and discussions

The relevant energy level we used is the $D_2$ line of $Rb^{87}$, as is shown in Fig. 1 (a). The experimental setup is shown in Fig. 2. Three single-mode tunable external cavity diode lasers (ECDL) (New Focus TLB-6900) are used as one probe and two coupling lasers. All the lasers have ∼300 kHz short-time linewidth (measured using beat frequency method). One probe and two coupling beams are co-propagating in the 75 mm long vapor cell to minimize the residual Doppler linewidth. A solenoidal coil surrounds the Rb vapor cell to supply a longitudinal magnetic field. A detailed description of the experimental setup can refer to [33].

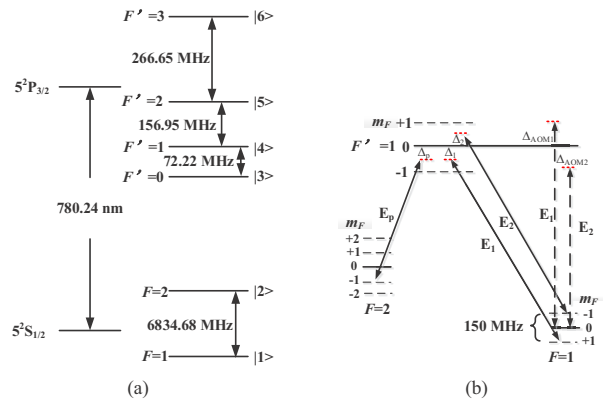

Fig. 1. (Color online) The relevant energy levels of $Rb^{87}$ for our experiment.

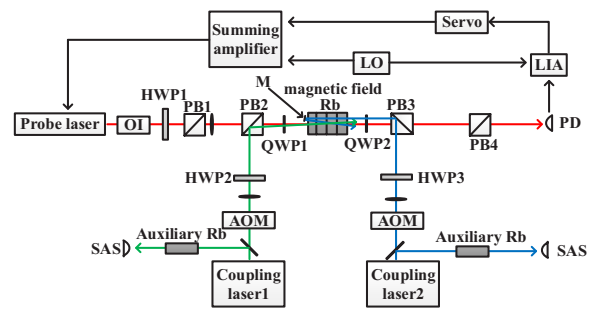

Fig. 2. (Color online) Schematic diagram of the experimental setup: PB1-PB4 (polarizing cubic beam splitters); HWP1-HWP3 (half-wave plates); QWP1, QWP2 (quarter-wave plates); AOM (acoustic optical modulator); PD (photodiode detector); OI (optical isolator); M (mini-reflective mirror reflecting coupling beam 2); LIA (Lock-in amplifier); LO (Local oscillator).

In the experiment, the probe beam's power is attenuated to below 50 μW to avoid the saturated absorption of the $Rb^{87}$ atoms and self-focusing effect via the half-wave plate 1 (HWP1) and polarized beam splitter 1 (PB1). Using two quarter-wave plates (QWP), we transform the polarization states of the laser beams $E_p$, $E_1$ and $E_2$ from linear to circular polarization when they enter into the Rb vapor cell. Adjusting the optic axis of the two QWPs appropriately, the probe and coupling beams will be seen in the atoms' frame as being $\sigma^+$ and $\sigma^-$ circularly polarized. The $\sigma^+$ polarized probe beam $E_p$ and coupling beam $E_2$ could only motivate the transition $\Delta m_F=+1$, while the $\sigma^-$ polarized coupling beam $E_1$ could only motivate the transition $\Delta m_F=-1$. We apply a magnetic field of 107.14 G in our experiment to make atomic levels split and the frequency separations between the sublevels of F=1 are 75 MHz. Thus, taking the transition rule and the polarization states of optical beams into consideration, a tripod-type EIT system for our experiment is formed. Since we will investigate the dependence of the EIT linewidth on the coupling fields' detuning, in the following, the coupling beams' frequency should be controlled precisely using the saturating absorption stabilization (SAS) method. About 10% of the coupling lasers' power are separated into auxiliary Rb cells for SAS measurement, while the detuning of the coupling fields are realized through two AOMs. As direct frequency modulation of the strong coupling fields should be avoided in order to reduce the residual amplitude modulation (RAM) noise in the laser fields [34], we tune the control voltage of external cavity length to scan the probe laser field's frequency.

As is discussed in our previous theoretical study [32], in a tripod configuration, an additional transition to the Lambda-type EIT system by another control field causes the occurrence of two distinct dark states. Interacting dark states makes one of the transmission peaks much narrower. By proper tuning of the coupling fields, the ultranarrow spectrum can be one order of magnitude narrower than that of a single dark state system. Now in Fig. 1 (b), when $E_1 > E_2$, the ultranarrow spectrum occurs in the EIT dip at $\Delta_p = \Delta_2$. In the experiment, the probe field is approximately $E_p$=48 $\mu$W and the two coupling fields are $E_1$=10 mW, $E_2$=0.49 mW respectively. Then, we control the modulation frequency of two AOMs at $\Delta_{AOM1}$=+50 MHz and $\Delta_{AOM2}$=-50 MHz. Taking the energy shift in the magnetic fields into consideration, the two coupling fields' detuning are $\Delta_1$=-25 MHz and $\Delta_2$=+25 MHz (detuning to level F'=1, $m_F$=0, as is shown in Fig. 1 (b)). As a result, two EIT dips occur when $\Delta_p$=-25 MHz and $\Delta_p$=+25 MHz, as is shown in Fig. 3. An ultranarrow linewidth (full width at half maximum) of 590 kHz is observed in the right EIT dip. This ultranarrow linewidth is about one order narrower than the natural linewidth of Rb[87]. The narrower EIT dip occurs where the relative frequency difference of probe and coupling beam 2 equals the energy level splitting of F=1 ($m_F$=-1) and F=2 ($m_F$=-1). Therefore, a good frequency offset locking effect of probe and coupling beam 2 via this reference signal could be expected.

In order to characterize the EIT signal, its dependence on the experimental parameters, such as the coupling fields' power and detuning are investigated. At first, we measure the linewidth of the narrower tripod-type EIT dip (the right dip in Fig. 3) under different powers of two coupling fields. In order to make sure the EIT effect is developed in the experiment, the power of two coupling beams are kept one order larger than the probe beam's. Figure 4 shows the variation of linewidth of the right EIT dip under different power of coupling beam 1 and coupling beam 2 as we keep the probe field $E_p$=48 $\mu$W. As is shown in Fig. 4(a), the narrower EIT dip linewidth is reduced dramatically as the power of coupling beam 1 increases to about 10 mW and then, the curve becomes gentle as the power further increases. While, as Fig. 4(b) shows, the narrower EIT dip becomes broaden as increasing the power of coupling beam 2. Also, we simulate this tripod-type EIT mechanism using density-matrix approach, which is similar to the description in our previous report [32]. For simplicity, we assume the relaxation rates from the exited state to each ground state are equal ($\gamma$=2$\pi$×2 MHz). The EIT dip profile is characterized under different power of two coupling beams. From these simulated EIT dips spectrum, we record the linewidth verse different coupling beams' power in Fig. 4 (c) and (d). As some factors in the simulation cannot be estimated precisely from the experimental process including the lasers' linewidth, the fluctuation of magnetic amplitude and the optical beams' diameter, there is a little deviation between the experimental and simulated result. While, the tendency of experimental and simulation result are shown in good agreement. So, in order to get an ultranarrow EIT dip, the power of coupling beam 1, coupling beam 2 are kept at about 10 mW, 0.49 mW, respectively, with a 48 $\mu$W probe beam in our experiment.

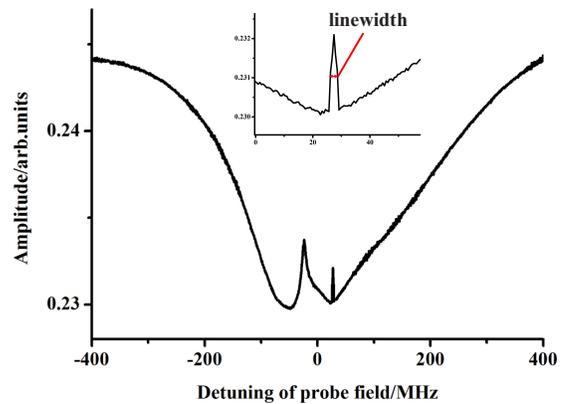

Fig. 3. (Color online) Two transparency windows induced by the double-dark resonances EIT system: the right window ($\Delta_p$=$\Delta_2$) is much narrower than the left one ($\Delta_p$=$\Delta_1$) as $E_2$=0.49 mW is weaker than $E_1$=10mW. (The x-axis is calibrated using SAS and tuning coefficient of probe laser.)

Secondly, the dependence of the narrower EIT dip's width on the coupling fields' detuning is studied. In our experiment, we record the linewidth of the narrower EIT dip (the right one in Fig. 3) as changing the two coupling fields' detuning with modulating the two AOMs. When the modulation frequency of the two AOMs are changed, we adjust the reflective angle of PB2 and PB3 to align the three laser beams. Two coupling fields' frequency detuning to F'=1 ($m_F$=0) from -50 MHz to +50 MHz with a step of 5 MHz is recorded in Fig. 5 (see the energy level in Fig. 1 (b)). It is clearly seen that the ultranarrow EIT spectrum exists in some regions (the green regions in Fig. 5), which corresponds to the case that the two EIT dips are not too far away or too closed. As is mentioned above, it means the ultranarrow spectrum occurs because of the interacting dark states under proper coupling fields' detuning. It would bring benefit to the application of offset locking the frequency difference of two lasers via the EIT signal, as the additional frequency stabilization of the coupling beams would not be necessary.

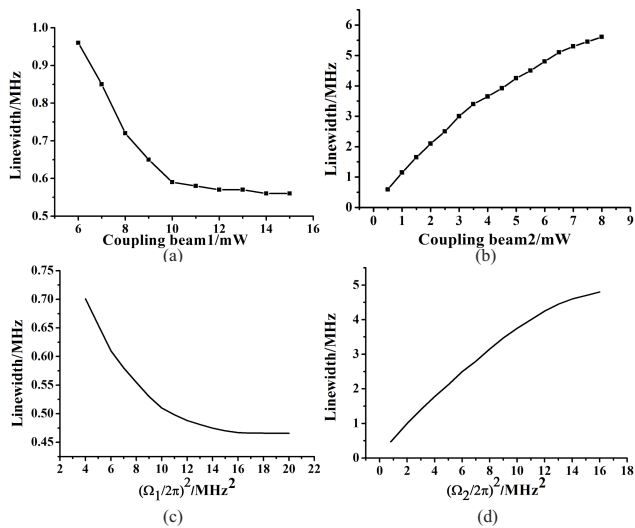

Fig. 4. Narrower EIT linewidth verse power of coupling beams:(a) $E_2$=0.49 mW; (b) $E_1$=10 mW; (c) $\Omega_p$=2π×0.1 MHz, $\Omega_2$=2π×0.9 MHz, $\Delta_1$=-2π×3 MHz, $\Delta_2$=+2π×3 MHz; (d) $\Omega_p$=2π×0.1 MHz, $\Omega_1$=2π×4 MHz, $\Delta_1$=-2π×3 MHz, $\Delta_2$=+2π×3 MHz;

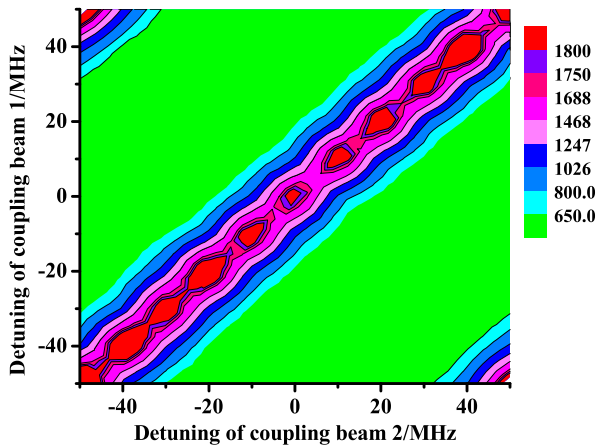

Fig. 5. (Color online) Narrower EIT linewidth verse coupling beams' detuning with $E_1$=10 mW and $E_2$=0.49 mW (Different color in the figure corresponds to different scale of EIT linewidth).

Then, we lock the frequency difference of the probe laser and the coupling laser 2 via two feedback channels, after the SAS setups, the auxiliary cells and the AOMs are removed. According to the experimental results in Fig. 4, 5, $E_1$, $E_2$ are kept at 10 mW and 0.49 mW with $\Delta_1$=-25 MHz and $\Delta_2$=+25 MHz (measured using a wavemeter). Thus, the ultranarrow EIT dip would appear in a wide coupling fields' detuning range even when the coupling fields' frequency is fluctuating. As is shown in Fig. 2, in our experimental process, a sinusoidal modulation voltage is applied from a local oscillator on the prober laser current to perform phase sensitive detection of the EIT signal with frequency of 12.5 MHz and the photediode signal is fed to a lock-in amplifier where the probe transmission is demodulated at the modulation frequency to derivate error signal. The error signal is plotted in the Fig. 6 which shows a ∼500 KHz frequency span between its two extremes and corresponds to the linewidth of the ultranarrow EIT dip. In order to reduce the short-time lasers' relative frequency fluctuation, a high speed feedback channel with bandwidth up to 1.5 MHz is used to control the diode injection current of probe laser. Another channel controls the external-cavity length of probe laser with bandwidth of 500 Hz, reducing long tome lasers' relative frequency fluctuation. The frequency stability recorded under different feedback conditions is shown in Fig. 7, which is converted from the error signal in Fig. 6. With two feed channels, the relative frequency fluctuation of two lasers is reduced from 1.5 MHz to 60 kHz in about 2000 s, which is narrower than the short-time linewidth of each individual laser. Also, the red signal in Fig. 7 shows that the short-time relative lasers frequency fluctuation cannot be reduced via a low frequency feedback channel and in this case, the relative frequency stability achieved is similar to the short-time linewidth of each individually laser, ∼300 kHz. In a similar experimental condition with two feedback channels, we also offset lock the relative frequency using a single dark resonance Lambda-type EIT system as we turn off the coupling laser 1 and keep the detuning $\Delta_2$ = 0 MHz. The result shows that the good offset locking effect retains in only about 3 minutes because the fluctuation of coupling laser 2's frequency. In Fig. 8, the experimental data is fast Fourier transform analyzed with the help of a digital storage oscilloscope to extract the lower part of the frequency noise power spectrum (<50 KHz) of the relative frequency of two lasers under different conditions. Both these noise power spectrum have some common mode components, including electrical line noise, low-frequency vibration noise and so on. At low frequency, both of frequency noise under one and two two feedback channels produce a similar performance. The frequency noise under two feedback channels keeps at a low value in a wide frequency range, while the one under one feedback channel reaches its maximum at about 1 KHz. It also concludes that the noise floor has been reduced apparently with the two feedback channels.

Finally, it is valuable to indicate that the current experimentally achieved laser frequency offset locking effect is limited by several factors. The dominant one at this stage is the power fluctuation of coupling lasers.

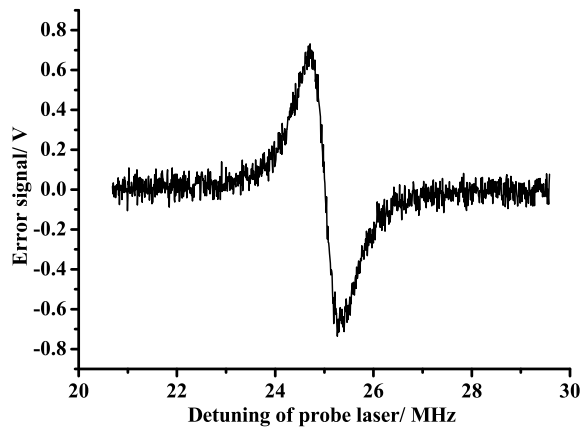

Fig. 6. Error signal of the ultranarrow EIT dip.

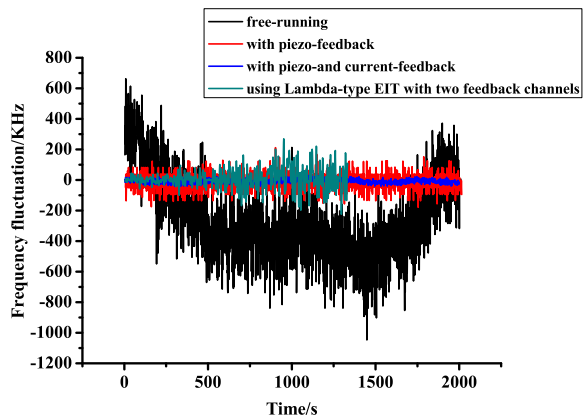

Fig. 7. (Color online) Relative frequency stability in a long time under different conditions.

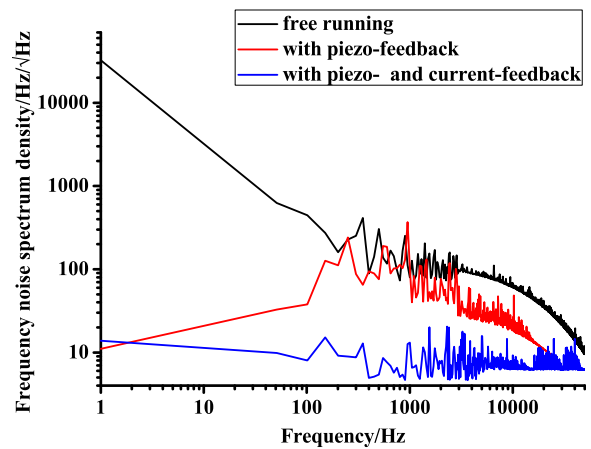

Fig. 8. (Color online) Frequency noise spectral density under different conditions.

From the experimental result in Fig. 4, the coupling lasers' power fluctuation have a great influence on the narrower EIT dip width for the tripod-type EIT system, which dramatically affects the offset locking effect. Other limiting factors include the fluctuation of magnetic amplitude and atomic temperature, the additional Zeeman splitting sublevels, the optical pumping effect of strong coupling field, the overlap area of three optical beams and the mechanical vibration of the experimental setup. With further experimental technical improvements, a better frequency offset locking effect can be expected.

### 3. Conclusion

In conclusion, we have offset locked the relative frequency of the probe laser and one coupling laser via the narrower EIT dip in the tripod-type double dark resonances system. The influence of laser fields' power and detuning on the tripod-type EIT profile has been detailed studied. In a wide coupling fields' detuning range, the ultranarrow spectrum with linewidth of ∼590 kHz has been obtained, which is about one order narrower than the natural linewidth of $Rb^{87}$. With low frequency feedback controlling the laser's external cavity-length and high frequency feedback controlling the diode current, the relative frequency of probe laser and one coupling laser has been offset locked. Without the additional frequency stabilization of coupling lasers, we have achieved the relative frequency fluctuation of 60 kHz in a long time ∼2000 s, which is narrower than the short-time linewidth of each individual laser.

### Acknowledgments

This work was supported by the National Natural Science Foundation of China (Grant Nos. 11274112, 91321101, 61108028 and 61178031) and Fundamental Research Funds for the Central Universities (Grant No. WM1313003).